\documentclass[12pt]{article} 
\usepackage[left= 0.8in, textwidth=6.9in, textheight=9.0in, top=1.0in]{geometry}
\usepackage{graphicx,caption}
\usepackage{amsmath,amssymb,verbatim}
\usepackage{enumitem}
\usepackage{color}  
\usepackage{bm}                                  

\newcommand{\MT}{\left[ \begin{array}{rrrrrrrrrrrrrrrrrrrr}}
\newcommand{\EM}{\end{array}\right]}
\newcommand{\EQ}{\begin{equation}\begin{alignedat}{4}} 
\newcommand{\EE}{\end{alignedat}\end{equation}}
\newcommand{\SEQ}{\begin{subequations}}
\newcommand{\ESE}{\end{subequations}} 

\newcommand{\Real}{\mathbb R}
\newcommand\norm[1]{\left\lVert#1\right\rVert}

\def\calH{{\cal H}}
\def\calY{{\cal Y}}

\def\bfp{{\bf p}}

\def\bfn{{\bf n}}
\def\bfx{{\bf x}}
\def\bfy{{\bf y}}

\def\bfu{{\bf u}}

\def\ds{\displaystyle}
\def\Fr{\ds \frac}

\title{\LARGE \bf
Feature-Informed Data Assimilation - \\ Definitions and Illustrative Examples
\thanks{This work was supported in part by AFOSR}
}

\author{Wei Kang\thanks{Department of Applied Mathematics, Naval Postgraduate School, Monterey, California, USA}, Daniel M. Tartakovsky\thanks{Department of Energy Science and Engineering, Stanford University, Stanford, California, USA},
Apoorv Srivastava\thanks{Stanford University, Stanford, California, USA}
}
\date{}
\begin{document}
\maketitle

\abstract{We introduce a mathematical formulation of feature-informed data assimilation (FIDA). In FIDA, the information about feature events, such as shock waves, level curves, wavefronts and peak value, in dynamical systems are used for the estimation of state variables and unknown parameters. The observation operator in FIDA is a set-valued functional, which is fundamentally different from the observation operators in conventional data assimilation. Demonstrated in three example, FIDA problems introduced in this note exist in a wide spectrum of applications in science and engineering.}

\section{Introduction}
Data assimilation (DA) is an interdisciplinary area in science that combines theories on estimation and stochastic systems with algorithms of scientific computing. It has a wide spectrum of applications in science and engineering, including numerical weather prediction, ocean forecasting, space weather, and the modeling and control of dynamical systems. In a DA process, the observations and the system's numerical model are combined to seek for an optimal estimation of the state of the system as it evolves in time. For readers who are interested in an overview of DA, there are many survey papers in the literature \cite{leeuwen2015,houtekamerzhang,bannister2017,fairbairn2014,law2012}. Most DA methods are based on Bayesian context. In numerical weather prediction, some widely used algorithms are four-dimensional variational data assimilation (4D-Var) \cite{bannister2017,fairbairn2014,rabier2005,xudaley2000}, ensemble Kalman filters (EnKF) \cite{houtekamerzhang,carvalho2018,kangxu2021}, and their varieties. There are also many engineering applications of DA such as \cite{SDC_FZ_IJIRA19} on robotics and \cite{sinsbecktartakovsky2015} on multi-fidelity modeling. 

The formulation of DA problems consists of a numerical model of the dynamical system and an observation operator. The system's model defines the underlying relationship of the state variables between time steps. The observation operator ensures that observations can be quantitatively compared with the state estimation. For a given system to be estimated, the format and accuracy of observations have to be taken into consideration in the development of DA algorithms. In numerical weather prediction, for instance, temperature, dew point temperature, pressure, wind direction and wind speed are some examples of observations. They are collected through various types of equipment such as balloons, radars and satellites. For observations whose operator is a function of the state variable at a known location, integrating the data into an assimilation algorithm is relatively straightforward. However, some observations may have a complicated operator whose evaluation depends on the environment in a region in space. For instance, this is typical for satellite observations in weather prediction. The assimilation of satellite data is a challenging problem \cite{eyre2020} because the observation operator is difficult to model or an existing operator's dimension is too high for DA. In other disciplines, DA is increasingly important nowadays due to the ever-growing complexity of dynamical systems in areas such as turbulent combustion and detonation, large-scale power systems, and pandemic propagation of infectious diseases. Many observations in these problems depend on the state over a region in space, not the value of state at individual points. For example, shock waves are a hallmark of detonation phenomena \cite{crane2019,crane2021}. The location of shocks can be observed by image sensors. However, this information cannot be modeled as a function of the system's state at fixed locations because shocks are a result of the global property of dynamical propagation. This type of observations based on phenomena, or features, of dynamical systems widely exist in science and engineering, such as combustion wavefront \cite{balasuriya2006} and peak-to-peak plot of chemical reactor \cite{piccardi2006,peng1990}. Information based on features cannot be formulated using conventional finite input functions. The need in a variety of  disciplines and applications to assimilate feature based information motivates us to introduce a general mathematical formulation of feature-informed DA (FIDA), and to explore essential mathematical concepts as well as effective computational tools for FIDA. 

In this note, we introduce a mathematical formulation of FIDA. In Section \ref{sec_2}, a conventional formulation of DA is briefly introduced as background. Then, a FIDA problem is formulated in which the observation operator is a set-valued functional in an infinite dimensional space. In Section \ref{sec_3}, the definition of FIDA is illustrated using three different application scenarios. Effective and efficient computational algorithms for this new problem of FIDA is a subject for future research.

\section{Problem formulation}
\label{sec_2}
In the formulation of a DA problem, it consists of a deterministic forward dynamical model, the probability model of the dynamical system model error, an observation operator, and the probability model of the observation noise. The goal of DA is to combine the observations, or sensor data, and the system's numerical model to seek for an optimal estimation of the state of the system as it evolves in time. It can also be applied to identify unknown parameters. 

\subsection{The formulation of conventional DA}
The problem and examples in this note are motivated by the DA of partial differential equations (PDEs). In a PDE, we denote $\bfu(t,\bfx,\bfp)$ a solution, $\bfx$ is the space variable, $t$ is time, $\bfp$ is a constant parameter whose value is either known or unknown. Using a first order PDE as an example, we have
\EQ
\label{sys_model_1a}
\bfu_t(t,\bfx,\bfp)&=F(t,\bfu,\bfu_\bfx,\bfp), 
\EE
This PDE can be approximated by systems of ordinary differential equations (ODEs) using various discretization methods. As an illustrative example, let $\{\bfx_1, \bfx_2,\cdots,\bfx_N\}$ be a set of grid points in space. Let 
\EQ
U(t,\bfp)=\MT \bfu(t,\bfx_1,\bfp)\\ \bfu(t,\bfx_2,\bfp)\\ \cdots \\ \bfu(t,\bfx_N,\bfp)\EM
\EE
be the discretized trajectory on grid points. It follows an ODE model,
\SEQ
\label{sys_model_1}
\begin{alignat}{4}
\Fr{dU}{dt}(t,\bfp)&=M(t,U(t,\bfp),\bfp) + w(t), &\quad& U\in\Real^n, \bfp \in \Real^{n_\bfp}\\
\bfy(t)&=H(U(t,\bfp))+v(t), &\quad&\bfy \in \Real^{n_\bfy}
\end{alignat}
\ESE
where $n, n_\bfp, n_\bfy$ are positive integers. The variable, $\bfy$, represents the observation (output of the system). The function, $H: \Real^n \rightarrow \Real^{n_\bfy}$, is the observation operator.  The solution $U(t,\bfp)$ is a differentiable function whose value represents the state of the system at $t$ on grid points, such as temperature, pressure, and flow speed in the case of an atmospheric model.  The model error and the sensor noise are represented by $w$ and $v$, respectively. {\it The goal of DA is to numerically estimate the value of $U(t,\bfp)$ as well as $\bfp$ by combining the observations with the system model (\ref{sys_model_1})}.

The observation operator is a function that represents the relationship of the observation and the state of the system. For example, if a sensor is placed at the gird point $\bfx_1$, then $H$ is a linear function in which
\EQ
H(U(t,\bfp))=HU(t,\bfp)\\
H=\MT I &0&\cdots&0\EM
\EE
where $I$ is the identity matrix whose dimension is the same as the length of $\bfu$. Of course, the measurement can take place at multiple points in space and the location may not be at grid points. In any case, $H$ is a function defined on $\Real^n$. 

\subsection{Feature-informed DA}
\label{sec_FIDAformulation}
For systems operating in extremely harsh environment, such as inside a rotating detonation engine (RDE), collecting data on the state of the system is very challenging, if not impossible. On the other hand, the location and propagation of some eye-catching phenomena, or feature events, such as shock waves, wavefront and peaking, can be observed. The location and magnitude of these feature events may contain valuable information that can be used to estimate the state and parameters in the system. The DA problem based on this type of information, or feature-informed DA, is formulated as follows,
\SEQ
\label{sys_model_2}
\begin{alignat}{4}
&\bfu_t(t,\bfx,\bfp)=F(t,\bfu,\bfu_\bfx,\bfp)+w(t), \;\;\bfu(t,\bfx,\bfp)\in\Real^n \label{sys_model_2a}\\
&\calH:  \bfu(t,\cdot,\bfp) \rightarrow \calY (t)=\{ \bfx^\ast | \mbox{ feature event takes place at } (t,\bfx^\ast)\}+v(t). \label{sys_model_2b}
\end{alignat}
\ESE
In (\ref{sys_model_2b}), $\calY(t)$ depends on the value of $\bfp$ although this is not reflected in the notation. A feature event can represent a variety of phenomena with different mathematical definitions.  For example, if a feature event represents a shock, or discontinuity, taking place at  $(t, x^\ast)$, then the observation operator is
\EQ
\label{eq_Hshock}
\calH:  \bfu(t,\cdot,\bfp) \rightarrow \calY (t)=\left\{ \bfx^\ast \left|  \norm{\bfu_\bfx(t,\bfx^\ast,\bfp)} > M \right.\right\}+v(t)
\EE
where $M>0$ is a large number. This observation operator can also represent a wavefront where a state variable changes significantly in space, such as combustion wavefronts \cite{lee1984} or moving weather fronts \cite{Freitag2013}. Feature events are not limited to locations where the gradient is large. For instance, the maximum value of a function is a feature event that can also provide information about the system. Suppose $\bfu$ is a scaler-valued function. If an observation can capture the points at which $\bfu$ has maximum value, we can define the observation operator by
\EQ
\label{eq_Hmax}
\calH:  \bfu(t,\cdot,\bfp) \rightarrow \calY (t)&=\arg\max_{\bfx} \bfu (t, \bfx, \bfp) +v(t)\\
& = \left\{ \bfx^\ast \left| \bfu(t,\bfx,\bfp)\leq \bfu(t,\bfx^\ast,\bfp) \mbox{ for all }\bfx \right. \right\}+v(t)
\EE
One may also include the maximum value as a piece of information for FIDA, 
\EQ
\label{eq_Hmax_2}
\calH:  \bfu(t,\cdot,\bfp) \rightarrow \calY (t)& = \left\{ (\bfx^\ast, \bfu^\ast) \left| \bfu(t,\bfx,\bfp)\leq \bfu(t,\bfx^\ast,\bfp) \mbox{ for all }\bfx \right. \right\}+v(t)
\EE
A level surface  of a function is the set of points in space at which the value of the function equals a constant. Level surfaces are often used to identify boundaries such as the invariant sets of dynamical trajectories or boundary layers in fluid mechanics. If the location of a level surface is observed, the observation operator for DA purpose can be defined by
\EQ
\label{eq_Hlevel}
\calH:  \bfu(t,\cdot,\bfp) \rightarrow \calY (t)&=\left\{ \bfx^\ast \left|  \bfu(t, \bfx^\ast,\bfp)=C  \right. \right\} +v(t)
\EE
where $C$ is a given constant. Note that (\ref{eq_Hshock}), (\ref{eq_Hmax}),  (\ref{eq_Hmax_2}) and (\ref{eq_Hlevel}) are simple examples to illustrate the idea. In general, wavefronts, maximum value and level surfaces may be characterized by functions other than $\bfu(t,\bfx,\bfp)$. The formula that defines an observation operator should be customized to follow the physics and mathematical rules in the application. 

\subsection{What are the differences?}
Comparing (\ref{sys_model_2}) to (\ref{sys_model_1}), the main difference lies in the observation operator. In (\ref{sys_model_1}), $H$ is a vector-valued function from $\Real^n$ to $\Real^{n_\bfy}$. The  FIDA formulated in (\ref{sys_model_2}) has a fundamentally different observation operator. Firstly, $\calH$ is not a function defined on a finite dimensional space. It is a functional, or a function of functions, defined on the space of integrable functions that contains the solutions, $\bfu(t,\cdot,\bfp)$, of the PDE. In the examples shown above, the value of observation is the solution of functional equations or differential inequalities, such as $ \norm{\bfu_\bfx(t,\bfx^\ast,\bfp)} > M$ or $\bfu(t,\bfx,\bfp)\leq \bfu(t,\bfx^\ast,\bfp)$. Secondly, the value of $\calH$ is not a vector of a fixed dimension. Its value is a set consisting of all points at which a feature event takes place, i.e., $\calH$ is a set-valued functional. For a fixed $t$, the number of points in $\calY (t)$ can be finite, such as finite shocks, or infinite, such as continuous wavefronts. At different time, the number of points in $\calY (t)$ can be different. For example, branches of shocks can merge as time evolves. As a result, the number of points in $\calY (t)$ decreases. 

These differences bring both opportunities and challenges to research. As a functional, $\calH$ carries ``global information" about $\bfu$, rather than local information of $\bfu$ near sensor locations. How does this property improve the overall observability/identifiability of the system? It is an open problem to study. An effective FIDA system can provide estimation for systems in harsh environment where sensor data is not available except for the propagation of observable phenomenon such as shock waves. Because the space of measurable functions is infinite dimensional, computational efficiency and the scalability of algorithms are essential. The curse-of-dimensionality is a challenge that must be addressed. Motivated by numerous successes of learning-based methods in the research on high dimensional computational PDEs, deep learning is a promising approach in the effort of overcoming the curse-of-dimensionality. Different from Kalman filter which is optimal for linear dynamical systems with Gaussian noise, shock wave occurs in nonlinear systems only. The optimality of estimation is very difficult to achieve. The study of FIDA calls for additional mathematical tools that are not used in conventional DA. For instance, the observation operator is a set-valued function. Its analysis and computation needs set-valued analysis \cite{aubin1990}. Overcoming the challenges and developing FIDA algorithms are long term research subjects that cannot be all covered in one article. The goal of this note is to introduce the mathematical problem formulation and to demonstrate by examples that feature-informed observation does provide valuable information for the purpose of estimating unknown variables/parameters in the system. 

\section{Application scenarios of FIDA}
\label{sec_3}
The formulation of FIDA in Section \ref{sec_FIDAformulation} covers a wide spectrum of applications. Three scenarios are introduced in the following that illustrate how information about feature events can be formulated in the form of (\ref{sys_model_2}).

\subsection{Detonation}
\begin{figure}[!ht]
\centering
\captionsetup{width=.8\linewidth}
\includegraphics[width = 4.5in]{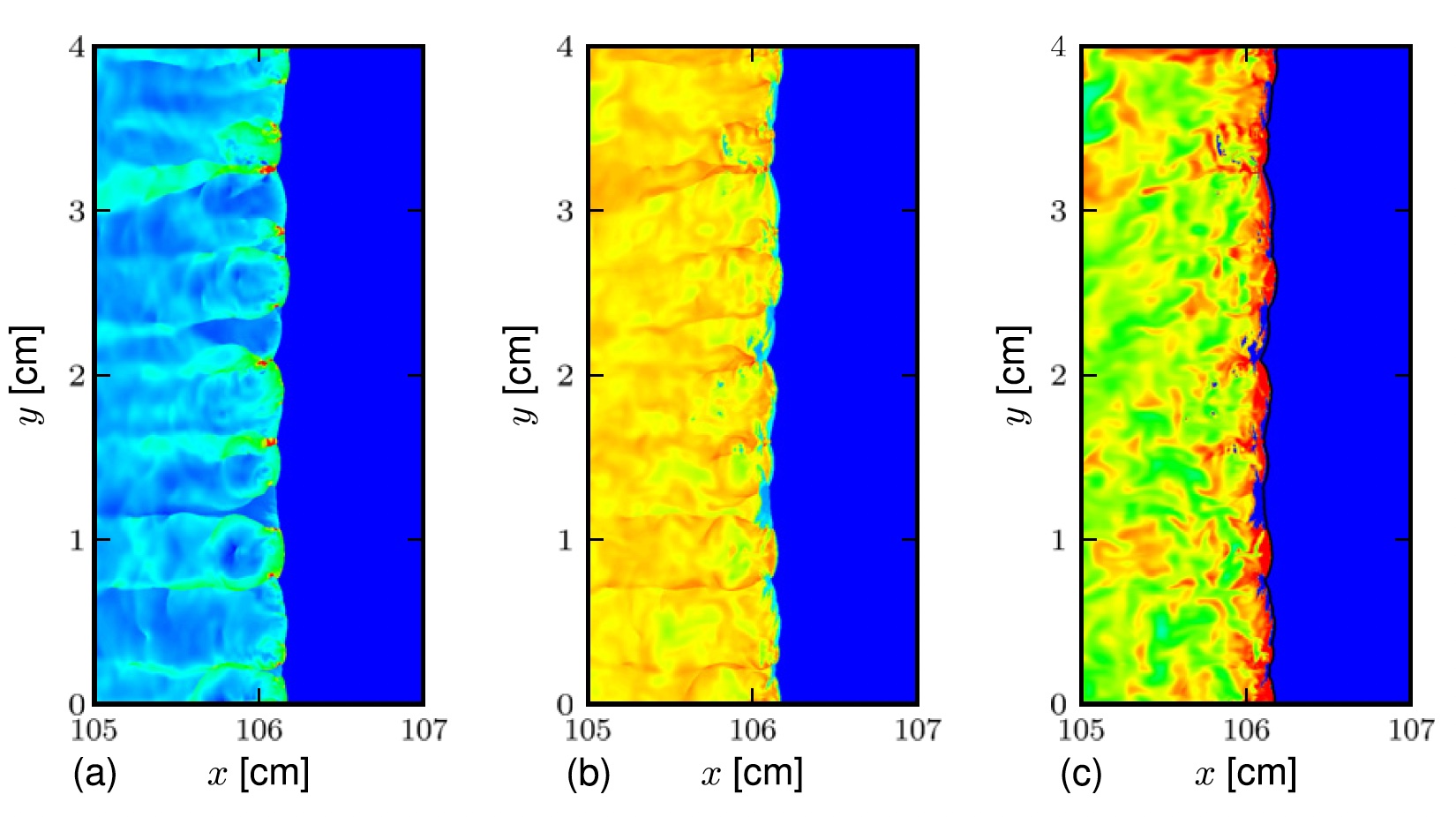}
\caption{An example of detonation. Figures courtesy of Taylor et al. \cite{Taylor_2013}: (a) pressure; (b) temperature; (c) OH mass fraction with shock surface indicated by black line. }
\label{fig_detonation}
\end{figure}

\begin{figure}[!ht]
\centering
\captionsetup{width=.8\linewidth}
\includegraphics[width = 4.5in]{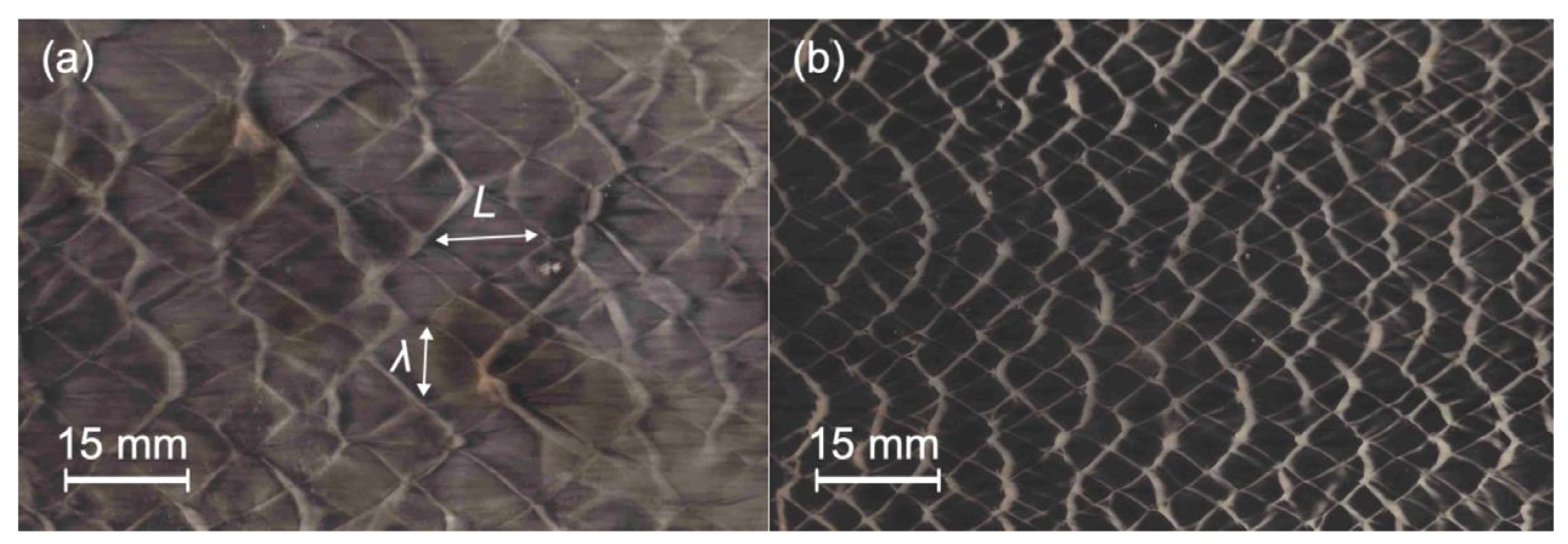}
\caption{An example of soot foil records. Figures courtesy of Crane et al. \cite{crane2019}: post-detonation soot traces with two different ozone concentrations}
\label{fig_sootfoil}
\end{figure}

Combustion wave propagation can occur at subsonic or supersonic velocities. The former is known as deflagration and the latter is called a detonation \cite{lee1984}. The exothermic front of a detonation propagates through a medium in the form of shock waves characterized by an abrupt, nearly discontinuous, change in pressure, temperature, and density of the medium. For example, the dynamics of detonation in a confined space follow a model that is based on the compressible Navier-Stokes equations with detailed chemistry. A simulation result in \cite{Taylor_2013} is shown in Figure \ref{fig_detonation}. When detonation propagates,  blasts and transverse waves have the maximum pressure. The kinematic history and the structure of detonation waves can be observed in experiments using sensing equipment such as high-speed cameras and soot foil \cite{crane2021,Taylor_2013,crane2019,koch_2020}. Two images of detonation experiments from \cite{crane2021} are shown in Figure \ref{fig_sootfoil}.

There are various ways of collecting data from shock waves. The observation operator should be defined accordingly. If the observation is the location of the wave front, the observation operator can be defined by the variation of a relevant variable, for instance,
\EQ
\calH:  \bfu(t,\cdot,\bfp) \rightarrow \calY (t)=\left\{ \bfx^\ast \left|  \norm{\mbox{gradient of pressure}} > M \right.\right\}+v(t)
\EE
for some $M>0$. Or if the observation is the location where the pressure is high, then 
\EQ
\calH:  \bfu(t,\cdot,\bfp) \rightarrow \calY (t)=\left\{ \bfx^\ast \left|  \mbox{pressure} > M \right.\right\}+v(t)
\EE

\subsection{Flame dynamics}
\begin{figure}[!ht]
\centering
\captionsetup{width=.8\linewidth}
\includegraphics[width = 4.5in]{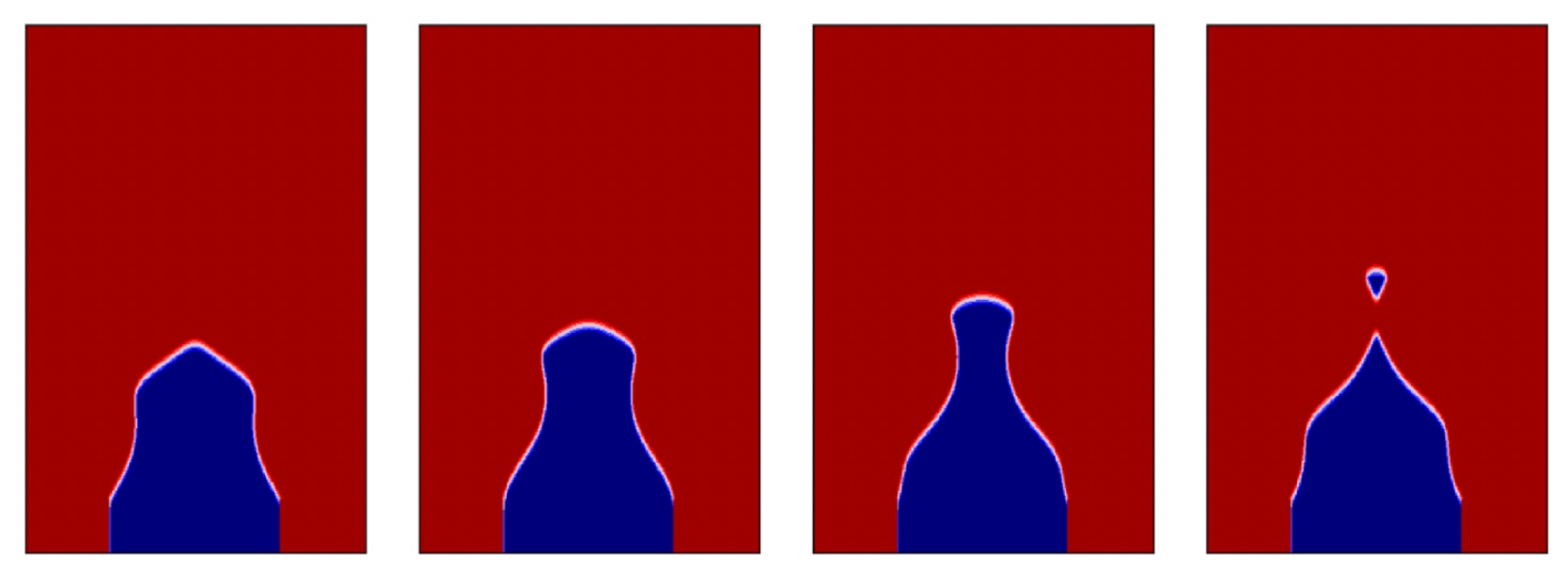}
\caption{A simulated example of  ducted premixed flame. Figures courtesy of Yu et al.  \cite{yu2019}. The fuel-air misture leaves the burner at the bottom of each frame. The infinitely thin flame surfaces separates the burnt (red) from the unburnt (blue) gas.}
\label{fig_flame}
\end{figure}

Shown in Figure \ref{fig_flame} is an example of ducted premixed flame studied in \cite{yu2019}.
The  flame surface is modeled and computed in \cite{yu2019} based on the following G-equation
\SEQ
\label{eq_flame}
\begin{alignat}{4}
&\Fr{\partial G}{\partial t}+(\bfu \cdot \bfn - s_L)=0\\
\mbox{subject to }& G(\bfx (t),t)-G(\bfx(0),0)=0 \label{eq_flame_b} \\
\mbox{and } & \sqrt{\nabla G\cdot \nabla G} - 1 =0
\end{alignat}
\ESE
where $\bfx$ is the position of a point on the flame surface, $\bfn$ is the normal vector at the point, $\bfu$ is the velocity field of the underlying medium, $s_L$ is the speed of the surface. The solution $G(\bfx,t)$ to the G-equation is called a generating function. The solution $\bfx (t)$ to (\ref{eq_flame_b}) gives the location of the flame surface at every time t. Note that (\ref{eq_flame_b}) implies that the flame surface is a level surface of the generating function. Suppose that the flame surface is observed for DA, then the observation operator is a functional
\EQ
\calH:  \bfu(t,\cdot,\bfp) \rightarrow \calY (t)=\left\{ \bfx^\ast \left|  G(\bfx^\ast,t)=C \right.\right\}+v(t)
\EE
where $C=G(\bfx(0),0)$ is a constant. The data of the level set provides valuable information that can be used to improve the
prediction of the dynamics of premixed flames. In \cite{yu2019}, data assimilation based on a EnKF to optimally calibrate the parameters and to quantify uncertainties in a model of flame dynamics. 

\subsection{Chaotic system estimation}
\begin{figure}[!ht]
\centering
\captionsetup{width=.8\linewidth}
\includegraphics[width = 3.0in]{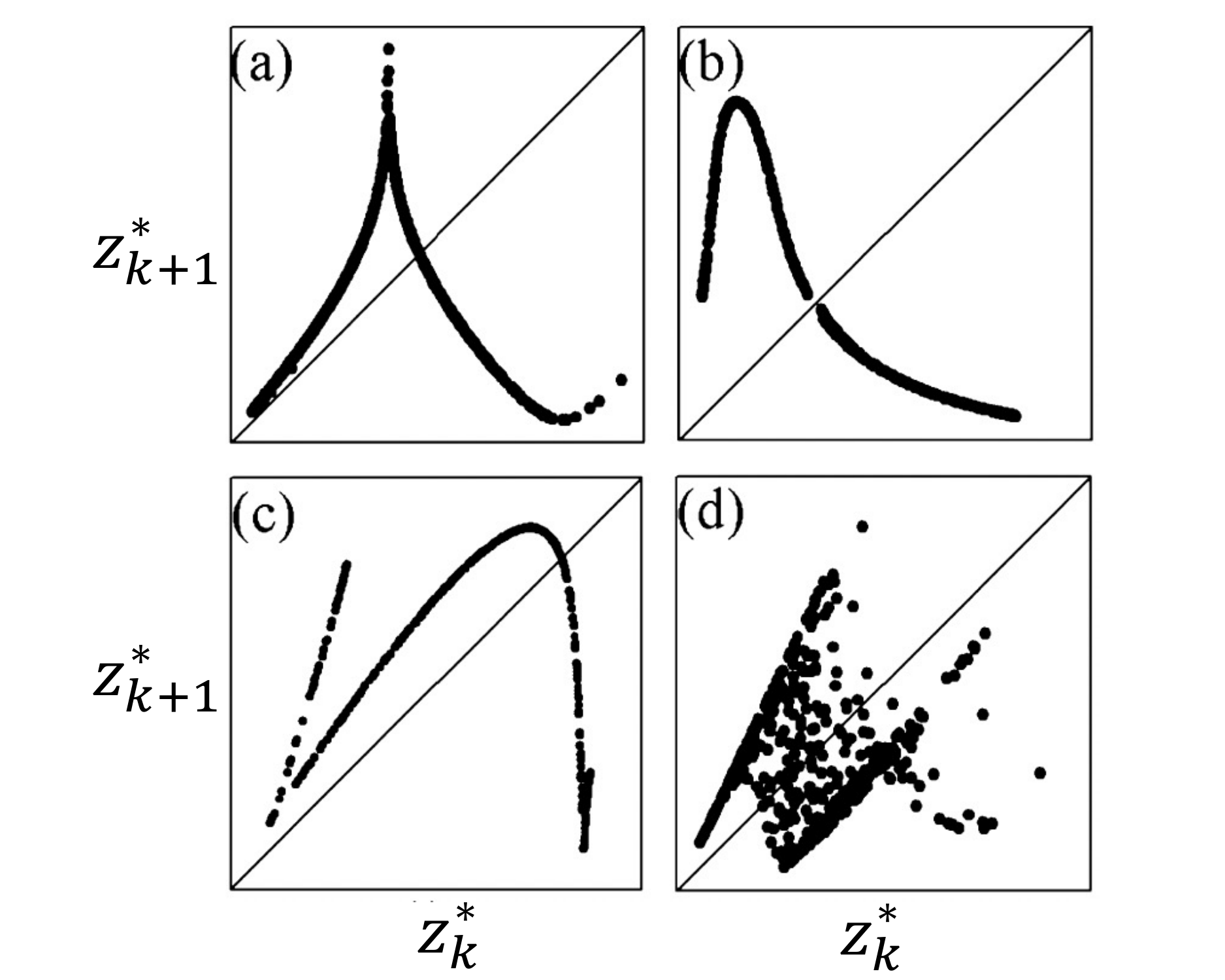}
\caption{Four examples of peak-to-peak plots. Figures courtesy of Piccardi \cite{piccardi2006}. (a) Lorenz system. (b) Chemical reactor. (c) Chua circuit. (d) R{\" o}ssler hyperchaotic system}
\label{fig_peak}
\end{figure}

Although the examples in previous sections are based on PDEs, the formulation of FIDA is applicable to ODEs as well.  Some feature events of ODEs can provide useful information for DA. The estimation of systems with chaotic trajectories has many applications in science and engineering such as the isothermal reaction in a thermodynamically closed system in a variety of chemical oscillations \cite{peng1990}. Among numerous approaches for parameter identification of chaotic systems, the one based on peak-to-peak dynamics \cite{piccardi2006} uses the information of local maximum value. Consider a system of ODEs
\EQ
\label{eq_ode_z}
\dot \bfx &= f(\bfx,\bfp)\\
z &= g(\bfx,\bfp)
\EE
where $\bfx$ is the state of the ODE system (not the space variable of a PDE), $\bfp$ represents unknown parameters to be identified. The scalar variable $z$ defines the feature used in estimation. Specifically, the peak value of $z$, or local maximum, can be observed. Along a trajectory of (\ref{eq_ode_z}), $z$ is a function of $t$. Suppose it achieves peak value at $t_k$, which is a finite or infinite sequence. Denote the peak value by $z_k$. This sequence is used in various applications to identify the unknown parameter $\bfp$ in the system model. The plots in Figure \ref{fig_peak} are four examples of peak-to-peak dynamics used for parameter identification \cite{piccardi2006} for (a) the Lorenz system \cite{lorenz1963}, (b) a chemical reactor \cite{peng1990}, (c) the Chua circuit  \cite{pivka1996} and (d) the R{\" o}ssler hyperchaotic system \cite{rossler1979}. 
They are all special cases of the FIDA problem formulated in  (\ref{sys_model_2}) with the following observation operator 
\EQ
\calH:  \bfx(t,\bfp) \rightarrow \calY (t)=&\left\{ (t^\ast, z^\ast) \left|  z^\ast = g(\bfx(t^\ast),\bfp), \right.\right. \\
&\left. \left. g(\bfx(t^\ast),\bfp) \geq z(\bfx(t),\bfp) \mbox{ for } t \mbox{ in a neighborhood of } t^\ast \right.\right\}+v(t)
\EE
For chaotic system with oscillations, $\calY(t)$ is a discrete sequence that can be ordered by the value of $t^\ast$. In \cite{piccardi2006}, a variational method is applied to this sequence to identify the unknown parameter in the model for several chaotic systems.  

\section{Conclusions}
In this note, we introduce a mathematical formulation for FIDA problems, the type of data assimilation that is based on the information of feature events. A FIDA problem consists of a dynamical system model and an observation operator. The main difference between conventional DA and FIDA problems lies in the observation operator in which the former is a function between finite dimensional spaces and the latter is a set-valued functional defined on a space of functions. Using examples we demonstrate that the formulation of FIDA is motivated by applications in a variety of areas in science and engineering. 

This note raises more questions than answers about the theoretical foundation and computational algorithms.  Does the data of feature events provide valuable information to improve the observability/identifiability of the system to be estimated? What are the advantages and disadvantages of DA algorithms, such as 4D-Var, EnKF, UKF and particle filters, for solving FIDA problems? How to efficiently compute the set-valued functional of observation operators? These open problems as well as engineering applications of FIDA are some of the focal topics in our future research. 

\vspace{0.1in}
\noindent
\textbf{Acknowledgement}. This work would not have been possible without the inspiring discussions with Dr. Chiping Li and Dr. Fariba Fahroo from AFOSR through several in person and virtual meetings. Dr. Li and Dr. Fahroo raised the question of DA based on the observed locations of shocks in combustion and detonation dynamics. The discussions revealed the facts that this type of observations is fundamentally different from conventional DA problems and developing a general mathematical formulation is important because feature-informed observations exist in a variety of applications. 

\bibliographystyle{ieeetr}
\bibliography{myreference}

\begin{thebibliography}{10}

\bibitem{leeuwen2015}
P.~J.~V. Leeuwen, Y.~Cheng, and S.~Reich, {\em Nonlinear Data Assimilation}.
\newblock Springer International, Switzerland, 2015.

\bibitem{houtekamerzhang}
P.~L. Houtekamer and F.~Zhang, ``Review of the ensemble {K}alman filter for
  atmospheric data assimilation,'' {\em Monthly Weather Review}, vol.~144,
  2016.

\bibitem{bannister2017}
R.~N. Bannister, ``A review of forecast error covariance statistics in
  atmospheric variational data assimilation. i: Characteristics and
  measurements of forecast error covariances,'' {\em Q. J. R. Meteorol. Soc.},
  vol.~134, pp.~1951--1970, 2008.

\bibitem{fairbairn2014}
D.~Fairbairn, S.~R. Pring, A.~C. Lorenc, and I.~Roulstone, ``A comparison of
  4{DV}ar with ensemble data assimilation methods,'' {\em Q. J. R. Meteorol.
  Soc.}, vol.~140, pp.~281 -- 294, 2014.

\bibitem{law2012}
K.~J. Law and A.~M. Stuart, ``Evaluating data assimilation algorithms,'' {\em
  Mon. Wea. Rev.}, vol.~140, pp.~3757--3782, 2012.

\bibitem{rabier2005}
F.~Rabier, ``Overview of global data assimilation developments in numerical
  weather-prediction centres,'' {\em Q. J. R. Meteorol. Soc.}, pp.~3215--3233,
  2005.

\bibitem{xudaley2000}
L.~Xu and R.~Daley, ``Towards a true 4-dimensional data assimilation algorithm:
  application of a cycling representer algorithm to a simple transport
  problem,'' {\em Tellus A: Dynamic Meteorology and Oceanography}, vol.~52,
  no.~2, pp.~109 -- 128, 2000.

\bibitem{carvalho2018}
S.~Vetra-Carvalho, P.~J. van Leeuwen, L.~Nerger, A.~Barth, M.~U. Altaf,
  P.~Brasseur, P.~Kirchgessner, and J.-M. Beckers, ``State-of-the-art
  stochastic data assimilation methods for high-dimensional non-gaussian
  problems,'' {\em Tellus A: Dynamic Meteorology and Oceanography}, vol.~70,
  no.~1, 2018.

\bibitem{kangxu2021}
W.~Kang and L.~Xu, ``Some quantitative characteristics of error covariance for
  {K}alman filters,'' {\em Tellus A: Dynamic Meteorology and Oceanography},
  vol.~73, no.~1, pp.~1--19, 2021.

\bibitem{SDC_FZ_IJIRA19}
D.~Chang, C.~R. Edwards, F.~Zhang, and J.~Sun, ``A data assimilation framework
  for data-driven flow models enabled by motion tomography,'' {\em
  International Journal of Intelligent Robotics and Applications}, vol.~3,
  no.~2, pp.~158--177, 2019.

\bibitem{sinsbecktartakovsky2015}
M.~Sinsbeck and D.~M. Tartakovsky, ``Impact of data asimilation on
  cost-accuracy tradeoff in multifidelity models,'' {\em SIAM/ASA J. Uncert.
  Quant.}, vol.~3, no.~1, pp.~954--968, 2015.

\bibitem{eyre2020}
J.~R. Eyre, S.~J. English, and M.~Forsythe, ``Assimilation of satellite data in
  numerical weather prediction. {P}art {I}: {T}he early years,'' {\em Q. J. R.
  Meteorol. Soc.}, vol.~146, pp.~49--68, 2020.

\bibitem{crane2019}
J.~Crane, X.~Shi, A.~V. Singh, Y.~Tao, and H.~Wang, ``Isolating the effect of
  induction length on detonation structure: hydrogen-oxygen detonation promoted
  by ozone,'' {\em Combustion and Flame}, pp.~44--52, 2019.

\bibitem{crane2021}
J.~Crane, X.~Shi, J.~T. Lipkowicz, A.~M. Kempf, and H.~Wang, ``Geometric
  modeling and analysis of detonation cellular stability,'' {\em Proceedings of
  the Combustion Institute}, no.~3, pp.~3585--3593, 2021.

\bibitem{balasuriya2006}
S.~Balasuriya, G.~Gottwald, J.~Hornibrook, and S.~Lafortune, ``High {L}ewis
  number combustion wavefronts: {A} perturbative {M}elnikov analysis,'' {\em
  {SIAM} Journal on Applied Mathematics}, vol.~67, no.~2, 2007.

\bibitem{piccardi2006}
C.~Piccardi, ``Parameter estimation for systems with low-dimensional chaos,''
  {\em {IFAC} Proceedings Volumes}, vol.~39, no.~8, pp.~291--296, 2006.

\bibitem{peng1990}
B.~Peng, S.~K. Scott, and K.~Showalter, ``Period doubling and chaos in a
  three-variable autocatalator,'' {\em J. Phys. Chem.}, pp.~5243--5246, 1990.

\bibitem{lee1984}
J.~H. Lee, ``Dynamic parameters of gaseous detonations,'' {\em Ann. Rev. Fluid
  Mech.}, vol.~16, pp.~3111--336, 1984.

\bibitem{Freitag2013}
M.~A. Freitag, N.~K. Nichols, and C.~J. Budd, ``Resolution of sharp fronts in
  the presence of model error in variational data assimilation,'' {\em Q. J. R.
  Meteorol. Soc.}, vol.~139, pp.~742--757, 2013.

\bibitem{aubin1990}
J.-P. Aubin and H.~Frankowska, {\em Set-Valued Analysis}.
\newblock Birkh{\"a}user, 1990.

\bibitem{Taylor_2013}
B.~D. Taylor, D.~A. Kessler, V.~N. Gamezo, and E.~S. Oran, ``Numerical
  simulations of hydrogen detonations with detailed chemical kinetics,'' {\em
  Proceedings of the Combustion Institute}, vol.~34, no.~2, pp.~2009--2016,
  2013.

\bibitem{koch_2020}
J.~Koch, M.~Kurosaka, and C.~Knowlen, ``Mode-locked rotating detonation waves:
  {E}xperiments and a model equation,'' {\em Physical Review E}, vol.~101,
  no.~013106, 2020.

\bibitem{yu2019}
H.~Yu, T.~Jaravel, M.~Ihme, M.~P. Juniper, and L.~Magri, ``Data assimilation
  and optimal calibration in nonlinear models of flame dynamics,'' {\em Journal
  of Engineering for Gas Turbines and Power}, vol.~141, no.~121010, 2019.

\bibitem{lorenz1963}
E.~N. Lorenz, ``Deterministic nonperiodic flow,'' {\em Journal of the
  Atmospheric Sciences}, vol.~20, pp.~130--141, 1963.

\bibitem{pivka1996}
L.~Pivka, C.~Wu, and A.~Huang, ``Lorenz equation and chua’s equation,'' {\em
  International Journal of Bifurcation and Chaos}, vol.~6, pp.~2443--2489,
  1996.

\bibitem{rossler1979}
O.~E. R{\" o}ssler, ``An equation for hyperchaos,'' {\em Physics Letters A},
  vol.~71, pp.~155--157, 1979.

\end{thebibliography}

\end{document}